\documentclass[longauth]{aa} 

\bibpunct{(}{)}{;}{a}{}{,} 
\usepackage{graphicx}
\usepackage{txfonts}
\begin{document}

   \title{MAGIC observations of the February 2014 flare of 1ES 1011+496 and ensuing constraint of the EBL density}

\author{
M.~L.~Ahnen\inst{1} \and
S.~Ansoldi\inst{2} \and
L.~A.~Antonelli\inst{3} \and
P.~Antoranz\inst{4} \and
A.~Babic\inst{5} \and
B.~Banerjee\inst{6} \and
P.~Bangale\inst{7,*} \and
U.~Barres de Almeida\inst{7,}\inst{26} \and
J.~A.~Barrio\inst{8} \and
J.~Becerra Gonz\'alez\inst{9,}\inst{27}
% ...
}

\institute { ETH Zurich, CH-8093 Zurich, Switzerland
\and Universit\`a di Udine, and INFN Trieste, I-33100 Udine, Italy
\and INAF National Institute for Astrophysics, I-00136 Rome, Italy
\and Universit\`a  di Siena, and INFN Pisa, I-53100 Siena, Italy
\and Croatian MAGIC Consortium, Rudjer Boskovic Institute, University of Rijeka, University of Split and University of Zagreb, Croatia
\and Saha Institute of Nuclear Physics, 1\textbackslash{}AF Bidhannagar, Salt Lake, Sector-1, Kolkata 700064, India
\and Max-Planck-Institut f\"ur Physik, D-80805 M\"unchen, Germany
\and Universidad Complutense, E-28040 Madrid, Spain
\and Inst. de Astrof\'isica de Canarias, E-38200 La Laguna, Tenerife, Spain; Universidad de La Laguna, Dpto. Astrof\'isica, E-38206 La Laguna, Tenerife, Spain
\and University of \L\'od\'z, PL-90236 Lodz, Poland
\and Deutsches Elektronen-Synchrotron (DESY), D-15738 Zeuthen, Germany
\and IFAE, Campus UAB, E-08193 Bellaterra, Spain
\and Universit\"at W\"urzburg, D-97074 W\"urzburg, Germany
\and Centro de Investigaciones Energ\'eticas, Medioambientales y Tecnol\'ogicas, E-28040 Madrid, Spain
\and Universit\`a di Padova and INFN, I-35131 Padova, Italy
\and Institute for Space Sciences (CSIC\textbackslash{}IEEC), E-08193 Barcelona, Spain
\and Technische Universit\"at Dortmund, D-44221 Dortmund, Germany
\and Unitat de F\'isica de les Radiacions, Departament de F\'isica, and CERES-IEEC, Universitat Aut\`onoma de Barcelona, E-08193 Bellaterra, Spain
\and Universitat de Barcelona, ICC, IEEC-UB, E-08028 Barcelona, Spain
\and Japanese MAGIC Consortium, ICRR, The University of Tokyo, Department of Physics and Hakubi Center, Kyoto University, Tokai University, The University of Tokushima, KEK, Japan
\and Finnish MAGIC Consortium, Tuorla Observatory, University of Turku and Department of Physics, University of Oulu, Finland
\and Inst. for Nucl. Research and Nucl. Energy, BG-1784 Sofia, Bulgaria
\and Universit\`a di Pisa, and INFN Pisa, I-56126 Pisa, Italy
\and ICREA and Institute for Space Sciences (CSIC\textbackslash{}IEEC), E-08193 Barcelona, Spain
\and Universit\`a dell'Insubria and INFN Milano Bicocca, Como, I-22100 Como, Italy
\and now at Centro Brasileiro de Pesquisas F\'isicas (CBPF\textbackslash{}MCTI), R. Dr. Xavier Sigaud, 150 - Urca, Rio de Janeiro - RJ, 22290-180, Brazil
\and now at NASA Goddard Space Flight Center, Greenbelt, MD 20771, USA and Department of Physics and Department of Astronomy, University of Maryland, College Park, MD 20742, USA
% ...
}

   \date{Received ; accepted }

 
  \abstract
   {In February-March 2014, the MAGIC telescopes observed the high-frequency peaked BL Lac 1ES 1011+496 (z=0.212) in flaring state at very-high energy  (VHE, E$>$100GeV). The flux reached a level more than 10 times higher than any previously recorded flaring state of the source.}
   {Description of the characteristics of the flare presenting the light curve and the spectral parameters of the night-wise spectra and the average spectrum of the whole period. From these data we aim at detecting the imprint of the Extragalactic Background Light (EBL) in the VHE spectrum of the source, in order to constrain its intensity in the optical band.}
   {We analyzed the gamma-ray data from the MAGIC telescopes using the standard MAGIC software for the production of the light curve and the spectra. For the constraining of the EBL we implement the method developed by the H.E.S.S. collaboration in which the intrinsic energy spectrum of the source is modeled with a simple function ($\leq 4$ parameters), and the EBL-induced optical depth is calculated using a template EBL model. The likelihood of the observed spectrum is then maximized, including a normalization factor for the EBL opacity among the free parameters.}
   {The collected data allowed us to describe the flux changes night by night and also to produce differential energy spectra for all nights of the observed period. The estimated intrinsic spectra of all the nights could be fitted by power-law functions. Evaluating the changes in the fit parameters we conclude that the spectral shape for most of the nights were compatible, regardless of the flux level, which enabled us to produce an average spectrum from which the EBL imprint could be constrained. The likelihood ratio test shows that the model with an EBL density $1.07$ (-0.20,+0.24)$_{stat+sys}$, relative to the one in the tested EBL template \citep{Dominguez2011}, is preferred at the $4.6$ $\sigma$ level to the no-EBL hypothesis, with the \textit{assumption} that the intrinsic source spectrum can be modeled as a log-parabola. This would translate into a constraint of the EBL density in the wavelength range [0.24 $\mu$m,4.25 $\mu$m], with a peak value at 1.4 $\mu$m of $\lambda F_{\lambda}=12.27_{-2.29}^{+2.75}$ nW m$^{-2}$ sr$^{-1}$, including systematics.}
   {}


   \maketitle
%

\section{Introduction} \label{intro}

1ES 1011+496 (RA: $10^{\rm h}\,15^{\rm m}\,04.1^{\rm s}$, DEC: $+49^{\circ}\,26^{\rm m}\,01^{\rm s}$) is an active galactic nucleus (AGN) classified as a high-frequency peaked BL Lac (HBL), located at redshift =0.212 \citep{MAGICdiscovery2007}. HBLs have spectral energy distributions (SED) characterized by two peaks, one located in the UV to soft X-ray band and the second located in the GeV to TeV range, which makes it possible to detect them in very-high-energy (VHE, E$>$100GeV) $\gamma$ rays. 1ES 1011+496 was discovered at VHE by the MAGIC Collaboration in 2007 following an optical high state reported by the Tuorla Blazar Monitoring Programme \citep{MAGICdiscovery2007}.

The observation of a bright source at intermediate redshift, like 1ES 1011+496, provides a good opportunity to constrain the impact of the Extragalactic Background Light (EBL) on the propagation of $\gamma$ rays over cosmological distances. The EBL is the diffuse radiation that comes from the contributions of all the light emitted by stars in the UV-optical and near infrared (NIR) bands. It also contains the infrared (IR) radiation emitted by dust after absorbing the starlight, plus a small contribution from AGNs \citep{HauserDwek2001}. VHE $\gamma$ rays from extragalactic sources interact with the EBL in the optical and NIR bands, producing electron-positron pairs, which causes an attenuation of the VHE photon flux measured at Earth \citep{Gould1967}.

Measuring directly the EBL is a challenging task due to the intense foreground light from interplanetary dust. For the optical band strict lower limits to the EBL have been derived from galaxy counts \citep{Madau2000,Fazio2004,Dole2006}. At NIR wavelengths, one way to access the EBL is by large-scale anisotropy measurements \citep[e.g.][]{Corray2004, Fernandez2010,Zemkov2014}. Making reasonable assumptions on the intrinsic VHE spectra of extragalactic sources (e.g. the limit in the hardness of the photon spectra of 1.5, coming from theoretical limits in the acceleration mechanisms), upper limits to the EBL density can be derived \citep[e.g.][]{Stecker1996,Aharonian2006,Mazin2007}. More recently, extrapolations of data from the Fermi Large Area Telescope have been used to set constraints to the intrinsic VHE spectra of distant sources, which, in combination with Imaging Atmospheric Cherenkov Telescopes (IACT) observations of the same objects, have also provided upper limits to the EBL density \citep{Georgan2010,Orr2011, Meyer2012}.

The Fermi collaboration employed a different technique to constrain the EBL density using a likelihood ratio test on LAT data from a number of extragalactic sources \citep{Fermi2012}. SEDs from 150 BL Lacs in the redshift range 0.03 - 1.6 were modeled as log parabolae in the optically-thin regime (E $<25$ GeV), then extrapolated to higher energies and compared with the actually observed photon fluxes. A likelihood ratio test was used to determine the best-fit scaling factor for the optical depth $\tau(E,z)$  according to a given EBL model, hence providing constraints of the EBL density relative to the model prediction. Several EBL models were tested using this technique \citep[e.g.][]{Stecker2006, Finke2010}, including the most widely and recently used by IACTs by \citet{Franceschini2008} and \citet{Dominguez2011}. They obtained a measurement of the UV component of the EBL of $3 \pm 1$ nW m$^{-2}$ sr$^{-1}$ at $z \approx 1$.

The H.E.S.S. collaboration used a similar likelihood ratio test to constrain the EBL taking advantage of their observations of distant sources at VHE  \citep{HESS2013}. The EBL absorption at VHE is expected to leave an imprint in the observed spectra, coming from a distinctive feature (an inflection point in the log flux--log E representation) between $\sim 100$ GeV and $\sim$5-10 TeV, a region observable by IACTs. This feature is due to a peak in the optical region of the EBL flux density, which is powered mainly by starlight. The H.E.S.S collaboration modeled the intrinsic spectra of several AGNs using simple functions (up to 4 parameters), then applied a flux suppression factor $\exp(- \alpha \times \tau(E,z))$, where $\tau$ is the optical depth according to a given EBL model and $\alpha$ a scaling factor. A scan over $\alpha$ was performed to achieve the best fit to the observed VHE spectra. The no-EBL hypothesis, $\alpha=0$, was excluded at the 8.8 $\sigma$ level,  and the EBL flux density was constrained in the wavelength range between 0.30 $\mu m$ and 17 $\mu m$ (optical to NIR) with a peak value of $15 \pm 2_{stat} \pm 3_{sys}$ nW m$^{-2}$ sr$^{-1}$ at 1.4 $\mu$m.

In \citet{Dominguez2013}, data from 1ES 1011+496 was used as part of a data set from several AGNs to measure the cosmic $\gamma$-ray horizon (CGRH). The CGRH is defined as the energy at which the optical depth of the photon-photon pair production becomes unity as function of energy. Using multi-wavelenght (MWL) data, \citeauthor{Dominguez2013} modeled the SED of each source, including 1ES 1011+496, doing a extrapolation to the VHE band. Then they made a comparison with the observed VHE data. In the case of 1ES 1011+496, they modeled the SED using the optical data from 2007 \citep{MAGICdiscovery2007} and X-ray data (from the \textit{X-Ray Timing Explorer}) taken in 2008 May, and compared it with the VHE data taken in 2007 by MAGIC. Their prediction was below the observed VHE data, which led to no optical-depth information. The prediction may have failed due to the lack of simultaneity in the data. A similar approach was presented by \citet{Nijil2010}, where they modeled the SED of PKS 2155-304 making a prediction for the VHE band and compared it to the observed data to give attenuation limits.

After the discovery of 1ES 1011+496 in 2007 \citep{MAGICdiscovery2007}, two more multi-wavelength campaigns have been organised by MAGIC: the first one between 2008 March and May \citep{Elina2015} and a second one divided in two periods, from 2011 March to April and 2012 from January to May \citep{MWLCornelia2014}. In all previous observations (including the discovery) the source did not show evidence of flux variability within the observed periods and the observed spectra could be fitted with simple power-law functions, with photon indices ranging between $3.2 \pm 0.4_{stat}$ and $4.0 \pm 0.5_{stat}$, and integral fluxes, above 200 GeV, between $(0.8 \pm 0.1_{stat}) \times 10^{-10}$ and $(1.6 \pm 0.3_{stat}) \times 10^{-11}$ photons cm$^{-2}$s$^{-1}$.

In this paper we present the analysis of the extraordinary flare of 1ES 1011+496 in 2014 February-March observed by the MAGIC telescopes, and apply a technique based on \citet{HESS2013} for constraining the EBL. The observations and the data reduction are described in Sect. \ref{sec:obs}, the results in Sect. \ref{sec:results}, the procedure for constraining the EBL in Sect. \ref{sec:emblems}, the inclusion of the systematic uncertainty is shown in Sect. \ref{sec:systunc}, and the results of the EBL constraint are discussed in Sect. \ref{sec:discu}.

\section{Observations \& Analysis} 
\label{sec:obs}

MAGIC is a stereoscopic system of two 17 m diameter Imaging Atmospheric Cherenkov Telescopes (IACT) situated at the Roque de los Muchachos, on the Canary island of La Palma (28.75$^{\circ}$N, 17.86$^{\circ}$W) at a height of 2200 m above sea level. Since the end of 2009, it has been working in stereoscopic mode with a trigger threshold of $\sim$50 GeV. During 2011 and 2012, MAGIC underwent a series of upgrades which results in a sensitivity of (0.66 $\pm$ 0.03)\% of the Crab nebula flux above 220 GeV in 50 hours at low zenith angles \citep{MAGICperformance20142,MAGICperformance2014}.

On February 5th 2014, VERITAS \citep{2002Weekes} issued an alert for the flaring state of 1ES 1011+496. MAGIC performed target of opportunity (ToO) observations for 17 nights during February-March 2014 in the zenith range of 20$^{\circ}-$56$^{\circ}$. After the quality cuts, 11.8 hrs of good data were used for further analysis. The data were taken in the so-called wobble-mode where the pointing direction alternates between four sky positions at 0.4$^{\circ}$ away from the source \citep{Fomin1994}. The four wobble positions are used in order to decrease the systematic uncertainties in the background estimation. The data were analyzed using the standard routines in the MAGIC software package for stereoscopic analysis, MARS \citep{Zanin2013}. 

\section{Results}
\label{sec:results}

After background suppression cuts, 6132 gamma-like excess events above an energy of 60 GeV were detected within $0.14^{\circ}$ of the direction of 1ES 1011+496. Three control regions with the same $\gamma$-ray acceptance as the ON-source region were used to estimate the residual background recorded together with the signal. The source was detected with a significance of $\sim 75$ $\sigma $, calculated according to \citet*[eq. 17]{LiMa}. 

 \begin{figure}
   \centering
   \includegraphics[width=\hsize]{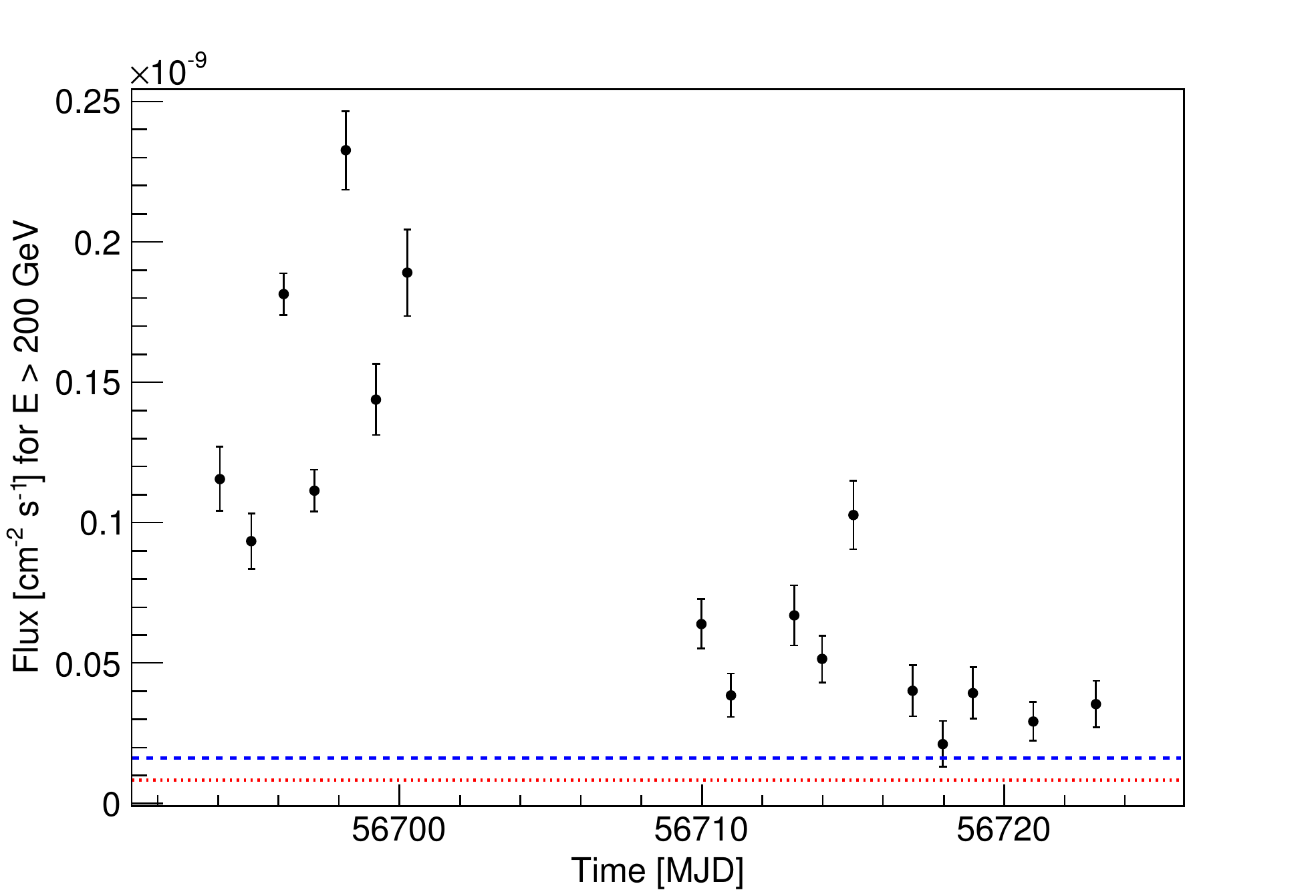}
      \caption{1ES 1011+496 light curve between February 6th and March 7th 2014 above an energy threshold of 200 GeV with a night-wise binning. The blue dashed line indicates the mean integral for the MAGIC observations of 2007 and the red dotted line the MWL campaign between 2011 and 2012.}
         \label{LC}
   \end{figure}

Fig. \ref{LC} shows the night by night $\gamma$-ray light curve for energies E $> 200$ GeV between February 6th and March 7th 2014. The emission in this period had a high night-to-night variability, reaching a maximum of $(2.3 \pm 0.1) \times 10^{-10}$ cm$^{-2}$s$^{-1}$, $\sim$14 times the mean integral flux measured by MAGIC in 2007 and 2008 for 1ES 1011+496  \citep{MAGICdiscovery2007,Riho2012} and $\sim$29 times the mean integral flux from the observation in 2011-2012 \citep{MWLCornelia2014}. For most of the nights the exposure time was $\sim$40 minutes, except for two nights (February 8th and 9th) in which the observations were extended to $\sim$2 hours. No significant intra-night variability was observed. The gap between observations seen in Fig. \ref{LC} was due to the strong moonlight period.

\begin{figure}
   \centering
   \includegraphics[width=\hsize]{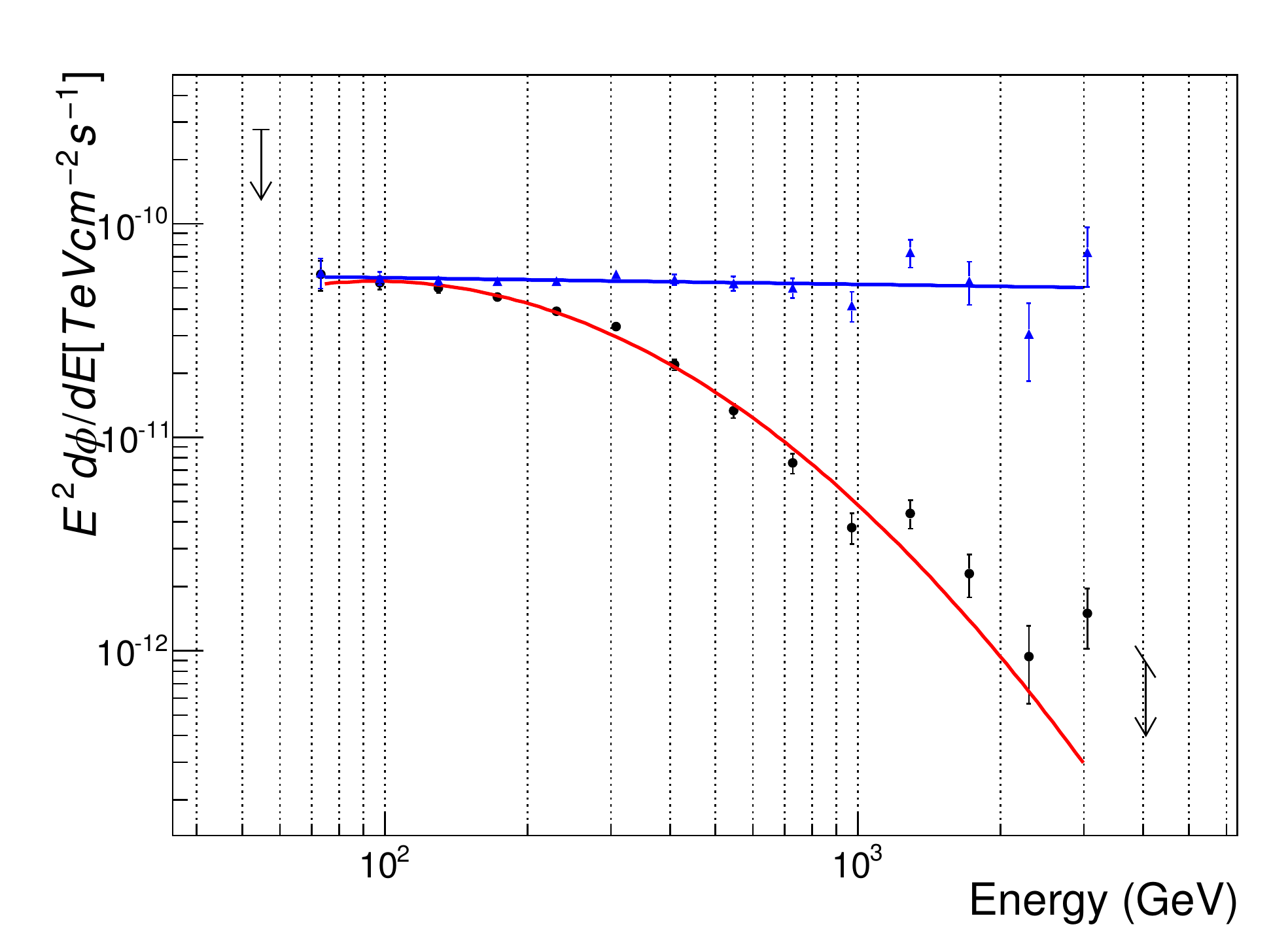}
      \caption{Spectral energy distribution (SED) of 1ES 1011+496 for the 17 nights of observations between February 6th and March 7th 2014. The black dots are the observed data and the blue triangles are the data after EBL de-absorption. The red line indicates the fit to a broken power-law with transition region function of the observed SED whereas the blue line indicates the fit to a power-law function of the de-absorbed SED. }
         \label{Spectrum}
   \end{figure}

The average observed spectral energy distribution (SED) is shown in Fig. \ref{Spectrum}. The estimated \textit{intrinsic} spectrum, assuming the EBL model by \citet{Dominguez2011}, can be fitted with a simple power-law function (PWL) with probability 0.35 ($\chi^{2}/$d.o.f.$=13.2/12$) and photon index $\Gamma=2.0 \pm 0.1$ and normalization factor at 250 GeV $f_{0}=(5.4 \pm 0.1) \times 10^{-11}$ cm$^{-2}$s$^{-1}$TeV$^{-1}$. The \textit{observed} spectrum is clearly curved. Several functions were tried to parametrize it: power-law with an exponential cut-off (EPWL), log-parabola (LP), log-parabola with exponential cut-off (ELP), power-law with a sub/super-exponential cutoff (SEPWL) and a smoothly-broken power-law (SBPWL). Of these, only the SBPWL,

\begin{equation}
\frac{dF}{dE}= f_{0} \left(\frac{E}{E_{0}}\right)^{-\Gamma_{1}}\left[1+ \left( \frac{E}{E_{b}} \right)^{g} \right]^{\frac{\Gamma_{2}-\Gamma_{1}}{g}}
\end{equation} 

 \noindent achieves an acceptable fit (P $= 0.17$, $\chi^{2}/$d.o.f.$=12.8/9$), though with a sharp change of photon index by $\Delta \Gamma= 1.35$ within less than a factor 2 in energy. For the SBPWL, the normalization factor at $E_{0}=250$ GeV is $f_{0}=(4.2 \pm 0.2) \times 10^{-11}$ cm$^{-2}$s$^{-1}$TeV$^{-1}$, the first index is $\Gamma_{1}=0.35 \pm 0.01$, the second index $\Gamma_{2}=1.7 \pm 0.1$, the energy break $E_{b}=298 \pm 21$ GeV and the parameter $g = 12.6 \pm 1.5$. Among the other, smoother functions, the next-best fit is provided by the LP (shown in Fig. \ref{Spectrum}), with P $= 1.7\times 10^{-3}$ ($\chi^{2}/$d.o.f.$=29.8/11$). The photon index for the LP is $\Gamma=2.8 \pm 0.1$, curvature index $\beta=1.0 \pm 0.1$ and normalization factor at $E_{0}=250$ GeV $f_{0}=(3.6 \pm 0.1) \times 10^{-11}$ cm$^{-2}$s$^{-1}$TeV$^{-1}$. This non-trivial shape of the observed spectrum, and its simplification when the expected effect of the EBL is corrected, strongly suggests this observation has high potential for setting EBL constraints.

\begin{figure}
   \centering
   \includegraphics[width=\hsize]{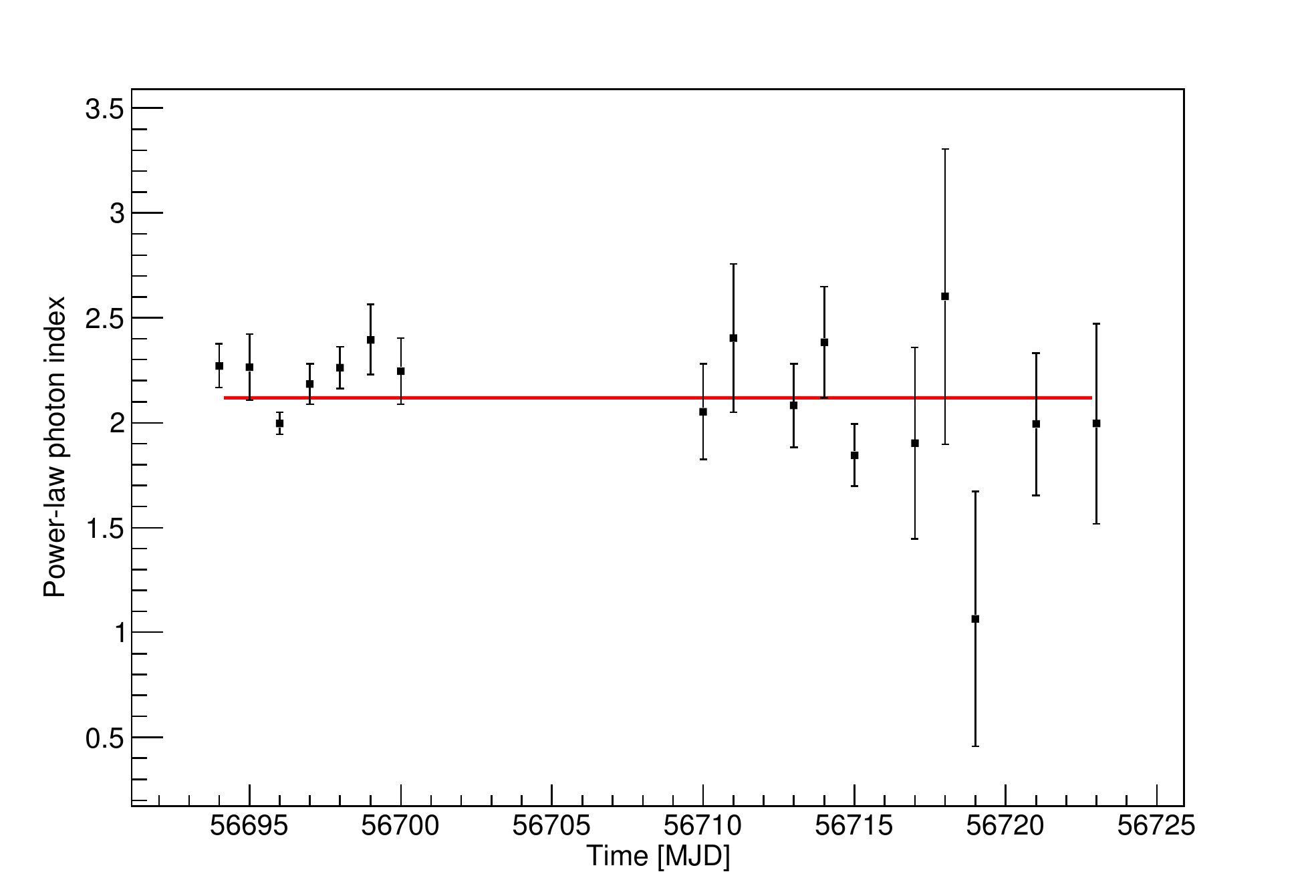}
      \caption{Evolution of the photon index from power-law fits to the de-absorbed night-wise spectra of 1ES 1011+496 between February 6th and March 7th 2014. The error bars are the parameter uncertainties from the fits. The red line represents the fit to a constant value, for which the probability is 10\%.} 
         \label{PLIndex}
   \end{figure}

The night-wise estimated intrinsic spectra could all be fitted with power-laws, and the evolution of the resulting photon indices is shown in Fig. \ref{PLIndex}. In the latter part of the observed period, the activity of the source was lower, resulting in larger uncertainties for the fits. There is no evidence for significant spectral variability in the period covered by MAGIC observations, despite the large variations in the absolute flux.

\section{EBL constraint}
\label{sec:emblems}

We follow the procedure described in \citet{HESS2013} for the likelihood ratio test. The absorption of the EBL is described as $e^{-\alpha \tau(E,z)}$ where $\tau(E,z)$ is the optical depth predicted by the model, which depends on the energy $E$ of the $\gamma$-rays and the redshift $z$ of the source. With the optical depth scaled by a factor $\alpha$, the observed spectrum is formed as:
\begin{equation}
\left(\frac{dF}{dE}\right)_{obs}=\left(\frac{dF}{dE}\right)_{int} \times \exp(- \alpha \times \tau(E,z))
\end{equation}

\noindent where $(dF/dE)_{int}$ is the intrinsic spectrum of the source. The emission of HBLs, like 1ES 1011+496, is often well described by basic synchrotron self-Compton (SSC) models \citep[e.g.][]{Tavecchio1998}. A population of electrons is accelerated to ultrarelativistic energies with a resulting power-law spectrum with index $\Gamma_{e}$ of about 2. The high energy electrons are cooled faster than the low energy ones, resulting in a steeper $\Gamma_{e}$. These electrons produce synchrotron radiation with a photon index $\Gamma=\frac{\Gamma_{e}+1}{2}=1.5$. In the Thomson regime the energy spectrum index of the inverse-Compton scattered photons is approximately the same as the synchrotron energy spectrum, whereas in the Klein-Nishima regime, the resulting photon index is even larger. These arguments put serious constraints to the photon index of the energy spectrum of VHE photons. Additionally, in most of the SSC models, the emission is assumed to be originated in a single compact region, which results in a smooth spectral energy distribution with two concave peaks. The shape of the individual peaks could be modified in a multizone model, where the emission is a superposition of several one-zone emission regions. However the general two-peak structure is conserved.

For the modeling of the intrinsic source spectrum we have used the same functions as in \citet{Mazin2007} and \citet{HESS2013} which were also used to fit the observed spectrum: PWL, LP, EPWL, ELP and SEPWL. We have added the additional constraint that the shapes cannot be convex, i.e. the hardness of the spectrum cannot increase with energy, as this is not expected in emission models, nor has it been observed in any BL Lac in the optically-thin regime. In particular, the un-absorbed part of BL Lac spectra measured by Fermi-LAT are well fitted by log-parabolas \citep{Fermi2012}.

The PWL and the LP are functions that are linear in their parameters in the log flux--log E representation (hence well-behaved in the fitting process), and both can model pretty well the de-absorbed spectrum found in Sect. \ref{sec:results}. The EPWL, ELP and SEPWL have additional (non-linear) parameters that are physically motivated, e.g. to account for possible internal absorption at the source. Note that these functions (except the PWL) can also mimic the \textit{overall} spectral curvature induced by the EBL over a wide range of redshifts, but will be unable to fit the inflection point (in the optical depth vs. log E curvature) that state-of-the-art EBL models predict around 1 TeV. We therefore expect an improvement of the fit quality as we approach the true value of the scaling factor $\alpha$, hence providing a constraint of the actual EBL density. The chosen spectral functions, however, do not exhaust \textit{all possible} concave shapes. Therefore the EBL constraints we will obtain are valid under the assumption that the true intrinsic spectrum can be well described (whitin the uncertainties of the recorded fluxes) by one of those functions. As we saw in section \ref{sec:results}, the 5-parameter SBPWL (not included among the possible spectral models) provides an acceptable fit to the \textit{observed} spectrum; if considered a plausible model for the intrinsic spectrum, it would severely weaken the lower EBL density constraint. On the contrary, the upper constraint (arguably the most interesting one VHE observations can contribute) from this work would be unaffected, as we will see below.

To search for the imprint of the EBL on the observed spectrum, a scan over $\alpha$ was computed, varying the value from 0 to 2.5. In each step of the scan, the model for the intrinsic spectrum was modified using the EBL model by \citet{Dominguez2011}, with the scaled optical depth using the expression (2) and then was passed through the response of the MAGIC telescopes (accounting for the effective area of the system, energy reconstruction, observation time). Then the Poissonian likelihood of the actual observation (the post-cuts number of recorded events vs E$_{est}$, in both the ON and OFF regions) was computed, after maximizing it in a parameter space which includes, besides the intrinsic spectral parameters, the Poisson parameters of the background in each bin of E$_{est}$ \footnote{Note that in the Poissonian likelihood approach we have included the point at $E\sim 55$ GeV which was shown just as an upper limit in Fig. \ref{Spectrum}, since it has an excess of just around 1 standard deviation above the background. The fit performed with the Poissonian likelihood approach has therefore one more degree of freedom than the $\chi^{2}$ fits reported in section \ref{sec:results}, and the 55 GeV point is included in Fig. \ref{Residuals}}. The maximum likelihood was thus obtained for each value of alpha. This likelihood shows a maximum at a value $\alpha=\alpha_{0}$, indicating the EBL opacity scaling which achieves a best fit to the data. A likelihood ratio test was then performed to compare the no-EBL hypothesis ($\alpha=0$) with the best-fit EBL hypothesis ($\alpha=\alpha_{0}$). The test statistics $TS=2\log (\mathcal{L}(\alpha=\alpha_{0}/\mathcal{L}(\alpha=0))$, according to Wilks theorem, asymptotically follows a $\chi^{2}$ distribution with one degree of freedom (since the two hypotheses differ by just one free parameter, $\alpha$).

Despite changing the flux level, the EBL determination method should work properly as long as the average intrinsic spectrum in the observation period can be described with one of the tested parameterizations. Assuming that is the case for the different states of the source, it will also hold for the average spectrum if the spectral \textit{shape} is stable through the flare. A simple way to check the stability of the spectral shape is fitting the points on Fig. \ref{PLIndex} to a constant value. The $\chi^{2}/$d.o.f. of this fit is $23.5/16$ and the probability is 10\%, so there is no clear signature of spectral variability ---beyond a weak hint of harder spectra in the second half of the observation period. A varying spectral shape would in any case need quite some fine tuning to reproduce, in the average spectrum, a feature like the one expected to be induced by the EBL.

 \begin{figure}
   \centering
   \includegraphics[width=\hsize]{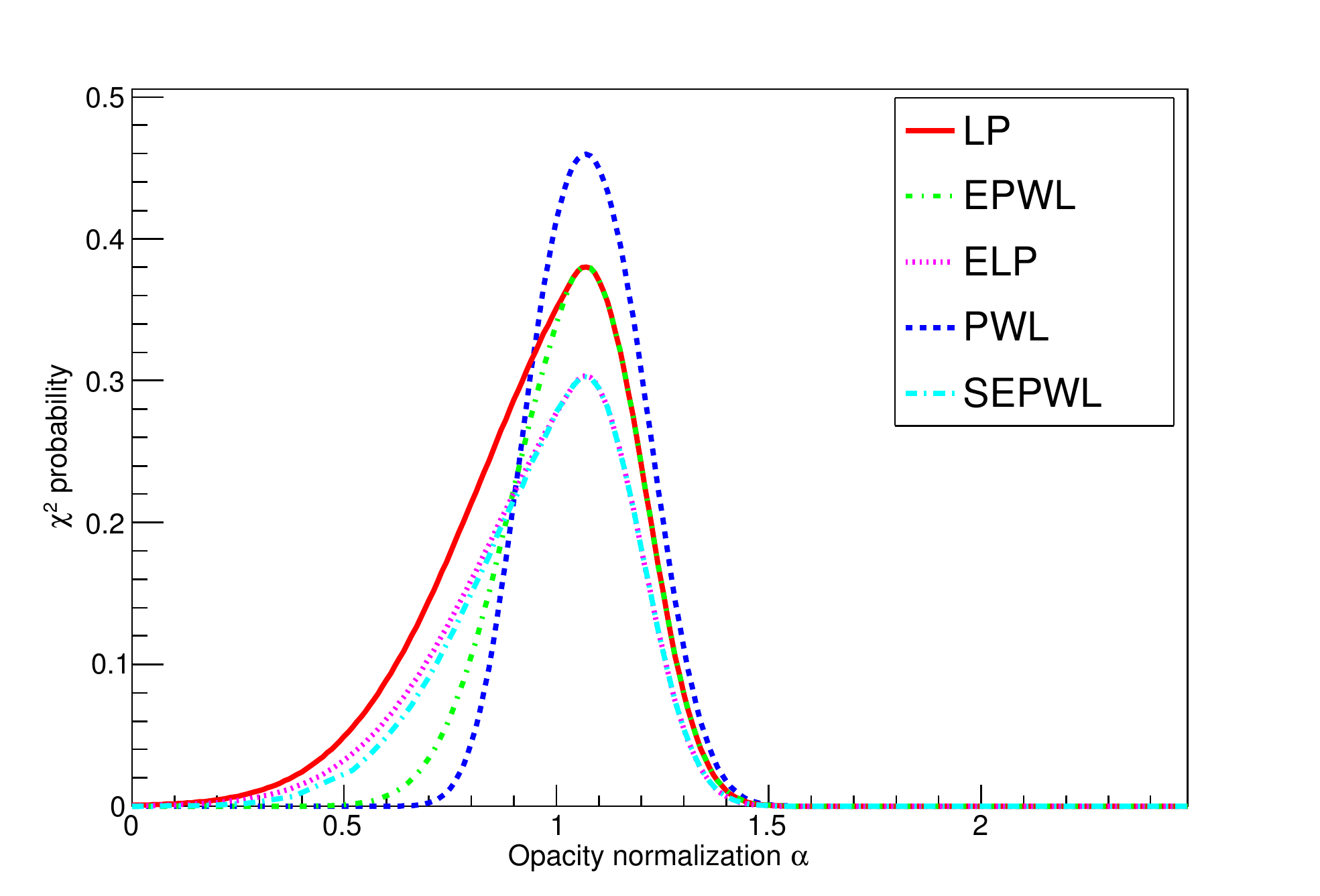}
      \caption{$\chi^{2}$ probability distributions for the average spectrum of the Feb-March flare of 1ES 1011+496 for the 5 models tested. PWL in blue (dashed line), LP in red (solid line), EPWL in green (dash-dot line), ELP in pink (dotted line) and SEPWL in light blue (long dash-dot line).}
         \label{PlotProbs}
   \end{figure}

 \begin{figure}
   \centering
   \includegraphics[width=\hsize]{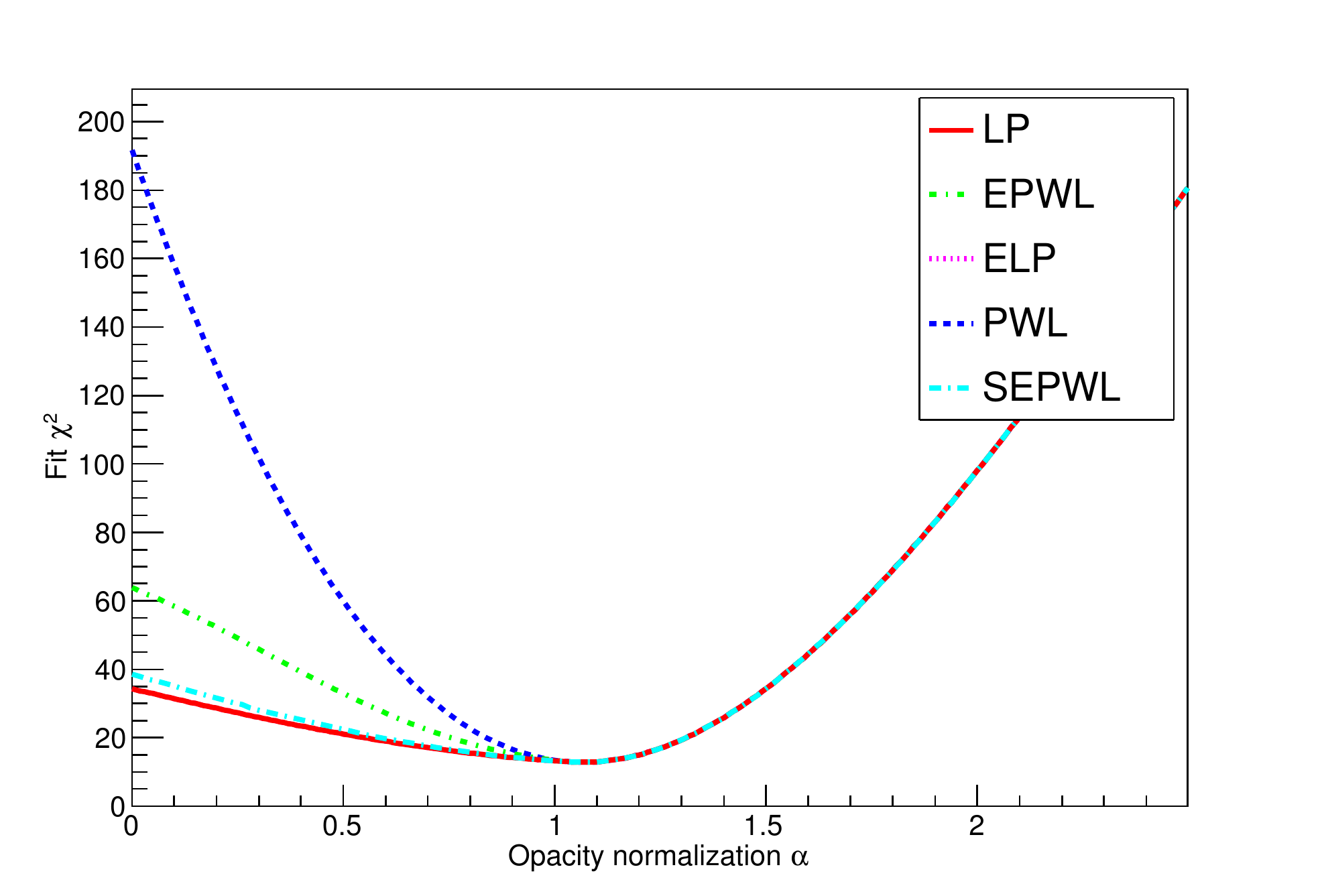}
      \caption{Fit $\chi^{2}$ distributions for the average spectrum of the Feb-March flare of 1ES 1011+496 for the 5 models tested. PWL in blue (dashed line), LP in red (solid line), EPWL in green (dot-dash line) and ELP in pink (dotted line) and SEPWL in light blue (long dash-dot line). The LP red line is overlapping ELP pink line. Notice how all curves converge after reaching the minimum.}
         \label{PlotChi2}
   \end{figure}

\begin{table}
\caption{$\chi^{2}$ probabilities (P) for the cases of $\alpha=0.0$ and $\alpha=1.07$}             
\label{table:Probas}      
\centering                          
\begin{tabular}{l c c}        
\hline\hline                 
Function & P($\alpha=0.0$) & P($\alpha=1.07$) \\    
\hline                        
   LP & $6.0 \times 10^{-4}$ & 0.38 \\      
   PWL & $7.0 \times 10^{-34}$ & 0.46 \\
   EPWL & $4.5 \times 10^{-9}$ &  0.38 \\
   ELP & $3.2 \times 10^{-4}$ & 0.30 \\
   SEPWL & $6.2 \times 10^{-5}$  &  0.30 \\ 
\hline                                   
\end{tabular}
\end{table}
 
Fig. \ref{PlotProbs} shows the $\chi^{2}$ probabilities for the five tested models, also listed in Table \ref{table:Probas} for the no-EBL case  ($\alpha=0.0$) and the best-fit $\alpha=1.07$. The model that gives the highest probability in the scanned range of $\alpha$ is the PWL. Following the approach in \citet{HESS2013} would lead us to choose the PWL as model for the intrinsic spectrum, as the next models in complexity (LP and EPWL) are not preferred at the 2 $\sigma$ level. However, choosing a PWL as the preferred model is rather questionable, since would not  allow any intrinsic spectral curvature, meaning that all curvature in the observed spectrum will be attributed to the EBL absorption. If this procedure is applied to a large number of spectra, as in \citet{Biteau2015ApJ}, individual $< 2$ $\sigma$ hints of intrinsic (concave) curvature might be overlooked and accumulate to produce a bias in the EBL estimation. In our case, the assumption of power-law intrinsic spectrum for 1ES 10111+496 would lead the likelihood ratio test to prefer the best-fit $\alpha$ value to the no-EBL hypothesis by as much as 13 $\sigma$. We prefer to adopt a more conservative approach, choosing the next-best function, the LP. Note however that at the best-fit $\alpha$, all the tested functions become simple power-laws, hence the fit probabilities at the peak depend only on the number of free parameters. At the best-fit $\alpha=1.07$, the parameters of the PWL are: photon index $\Gamma=1.9 \pm 0.1$ and normalization factor at 250 GeV $f_{0}=(9.2 \pm 0.2) \times 10^{-10}$ cm$^{-2}$s$^{-1}$TeV$^{-1}$. The other functions have the same values for these parameters.

Going deeper in the behaviour of the fits for the five models, it can be seen in the Fig. \ref{PlotChi2} that after reaching the minimum, the $\chi^{2}$ are identical for all models. This happens because of the concavity restriction imposed to the functions. After reaching the point where the EBL de-absorption results in a straight power-law intrinsic spectrum, all three functions converge, and the de-absorbed spectra becomes more and more convex as $\alpha$ increases (so no concave function can fit it any better than a simple power-law). The shape of the spectrum observed by MAGIC is thus very convenient for setting upper bounds to the EBL density, under the adopted assumption that convex spectra are ``unphysical''.

 \begin{figure}
   \centering
   \includegraphics[width=\hsize]{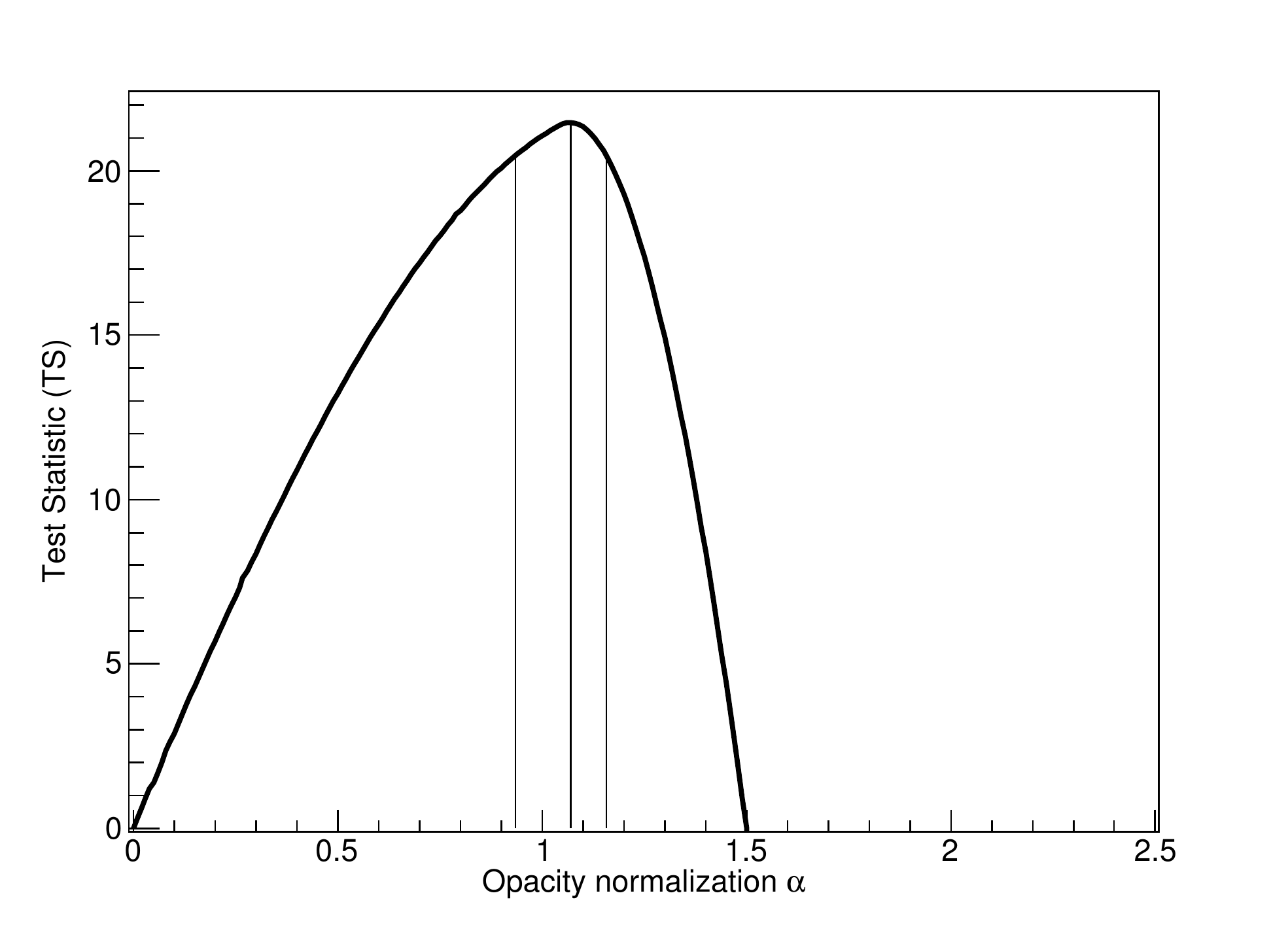}
      \caption{Test statistics distribution for the data sample for the 2014 Feb-March flare of 1ES 1011+496. The vertical lines mark the maximum and the uncertainty corresponding to 1 $\sigma$.}
         \label{TSallnights}
   \end{figure}

Given the arguments in the previous lines, we take the LP as our model for the intrinsic spectrum. For the data sample from the 2014 February-March flare of 1ES 1011+496, the test statistics has a maximum of TS = 21.5 at $\alpha_{0}=1.07_{-0.13}^{+0.09}$ (Fig. \ref{TSallnights}). This means that the EBL optical depth from the model of \citet{Dominguez2011} scaled by the normalization factor $\alpha_{0}$ is preferred over the null EBL hypothesis with a significance of 4.6 $\sigma$. Using the EBL model of \citet{Franceschini2008} as template (as in \citet{HESS2013}) the test statistic using the LP as model for the intrinsic spectrum has a maximum of TS=20.6 at $\alpha_{0}=1.14_{-0.14}^{+0.09}$, which is compatible with the result using \citet{Dominguez2011} within statistical uncertainty.

We again remark that allowing for other concave spectral shapes, like the SBPWL, would severely affect our lower EBL bound. This would also be the case for earlier published EBL lower constraints based on gamma-ray data ---especially those in which the PWL is among the allowed models for the intrinsic spectrum. For the observations discussed in the present paper, the SBPWL would achieve an acceptable fit even in the no-EBL assumption. This and earlier constraints on the EBL density through its imprint on $\gamma$-ray spectra hence rely on somewhat tentative assumptions on the intrinsic spectra ---but assumptions which, as far as we know, are not falsified by any observational data available on BL Lacs. On the other hand, the upper bound we have obtained is robust, since it is driven by the fact that convex spectral shapes (completely unexpected for BL Lacs at VHE) would be needed to fit our observations if EBL densities above the best-fit value are assumed.

\section{Systematic uncertainty}
\label{sec:systunc}

The MAGIC telescopes has a systematic uncertainty in the absolute energy scale of 15\% \citep{MAGICperformance2014}. The main source of this uncertainty is the imprecise knowledge of the atmospheric transmission. In order to assess how this uncertainty affects the EBL constraint, the calibration constants used to convert the pixel-wise digitized signals into photoelectrons were multiplied by a scaling factor (the same for both telescopes) spanning the range -15\% to +15\% in steps of 5\%. This procedure is similar as the one presented by \citet{MAGICperformance2014}.

For each of the scaling factors the data were processed in an identical manner through the full analysis chain, starting from the image cleaning, and using in all cases the standard MAGIC MC for this observation period. In this way we try to asses the effect of a potential miscalibration between the data and the MC simulation. 

For all scaled data samples, $\chi^{2}$ profiles for $\alpha$ between 0 and 2.5 were computed. From the 1 $\sigma$ uncertainty ranges in $\alpha$ obtained for the different shifts in the light scale, we determine the largest departures from our best-fit value $\alpha_{0}$, arriving to the final result $\alpha_{0}=1.07$ (-0.20,+0.24)$_{stat+sys}$.
  
\section{Discussion}
\label{sec:discu}

\begin{figure}
   \centering
   \includegraphics[width=\hsize]{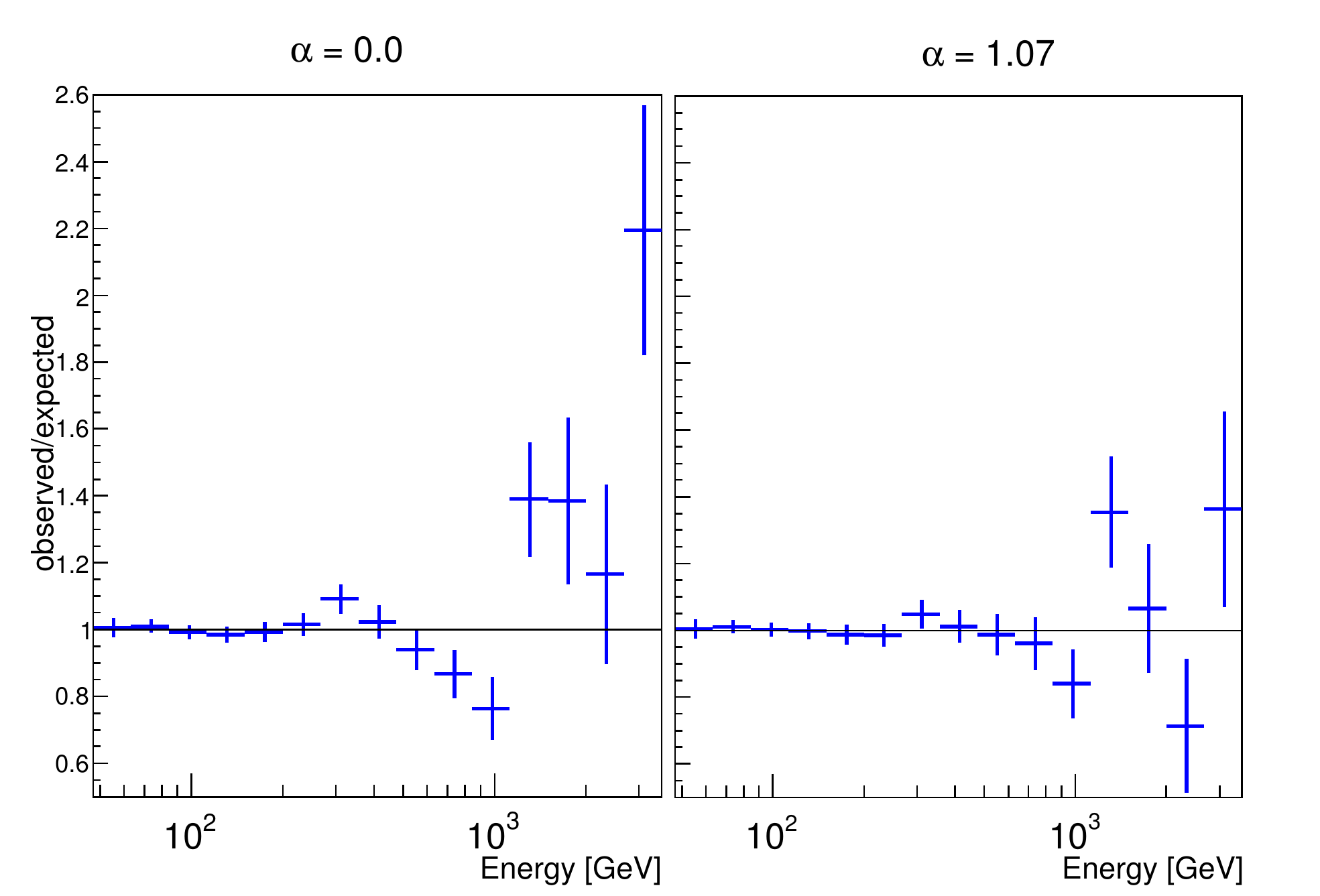}
      \caption{Ratio between the observed events and the expected events from the model of the intrinsic spectrum for two normalization values of the EBL optical depth, $\alpha=0$ to the left and $\alpha=1.07$ to the right, which corresponds to the normalization where the maximum TS was found. In both plots the line corresponding to a ratio=1 is shown.}
         \label{Residuals}
   \end{figure}

The relation of the $\gamma$-ray of energy $E_{\gamma}$ from the source (measured in the observed frame) and the EBL wavelength at the peak of the cross section for the photon-photon interaction is given by:

\begin{equation}
\lambda_{EBL}(\mu \text{m}) = 1.187 \times E_{\gamma}(\text{TeV}) \times (1+z)^{2}
\end{equation}

\noindent where $z$ is equal or less than the redshift of the source. The energy range used for our calculations was between 0.06 and 3.5 TeV. However, the constraining of the EBL following the method from \citet{HESS2013} is based in the fact that after de-absorbing the EBL effect, with the right normalization, the feature between $\sim100$ GeV and $\sim$ 5-10 TeV is suppressed. In Fig. \ref{Residuals} we show a comparison between two cases where the residuals were computed (ratio between the observed events and the expected events from the model). The plot on the left shows the residuals for the null EBL hypothesis $\alpha=0$, while the right pad shows the same plot for the case of the best fit EBL scaling $\alpha=1.07$. The differences start to show after 200 GeV, a region where the EBL introduces a feature (an inflection point) that cannot be fitted by the log-parabola. This is the feature that drives the TS value on which the EBL constraint is based. We therefore calculate the EBL wavelength range for which our conclusion is valid from the VHE range between 0.2 and 3.5 TeV.

\begin{figure*}
   \centering	
            \includegraphics[width=15cm]{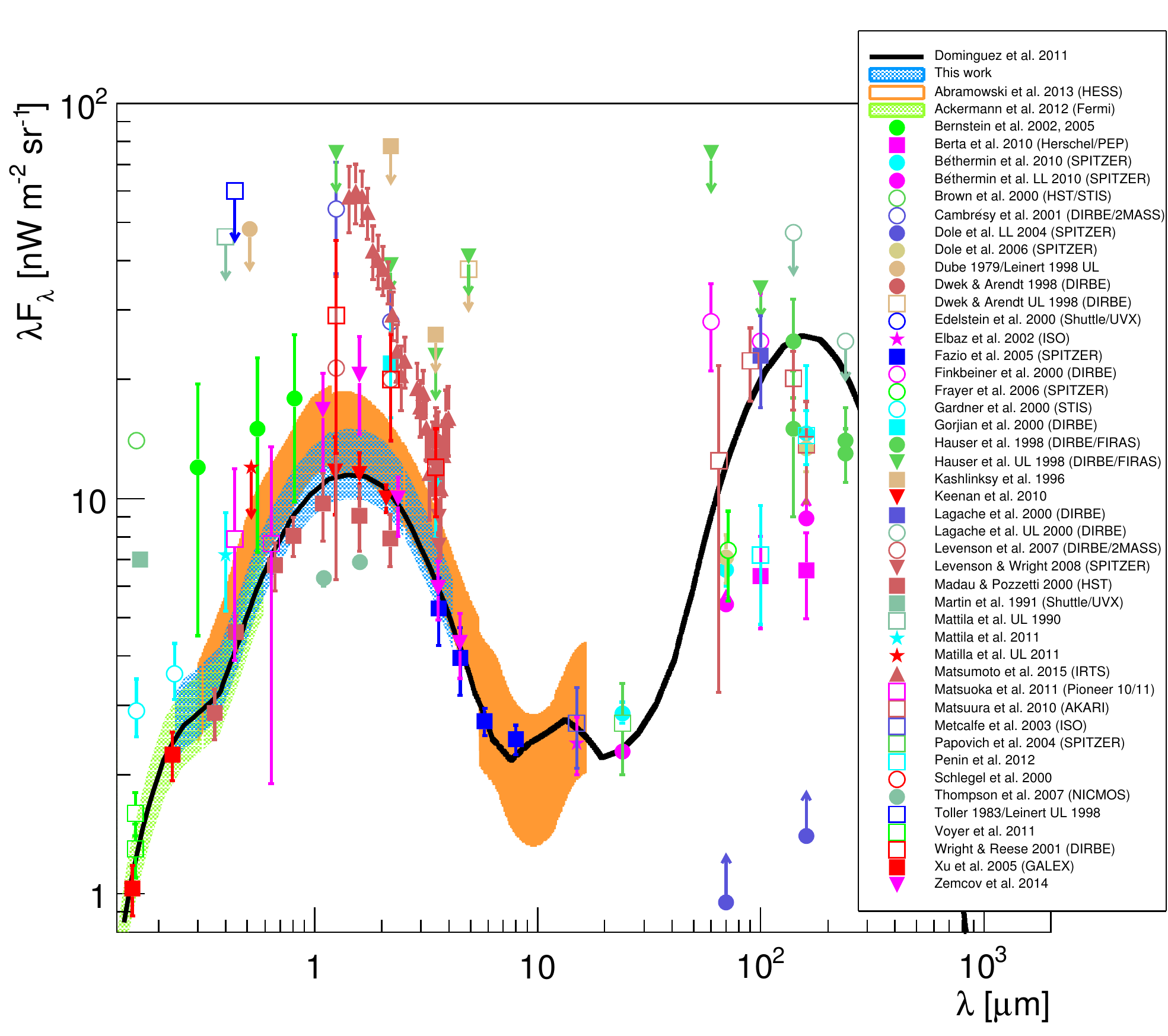}
      \caption{Extragalactic background light intensity versus wavelength at $z=0$. The solid black line is the EBL template model \citep{Dominguez2011} that we used for our calculations. The azure shaded area spans the wavelength range for which our constraint is valid, scaled from the EBL template. The width of the shaded area includes the statistical and systematic uncertainties. The orange area is the EBL constraint by \citet{HESS2013} and the green shaded area is the constraint by \citet{Fermi2012}. As a comparison we include the direct measurements by \citet{Dwek1998}, \citet{Hauser1998},  \citet{Finkbeiner2000}, \citet{Lagache2000}, \citet{Gorjian2000}, \citet{Cambresy2001}, \citet{Wright2001}, \citet{Bernstein2005},  \citet{Matsumoto2005}, \citet{Matsuoka2011}, and \citet{Matsumoto2015}. Also galaxy-count data is included, from \citet{Brown2000}, \citet{Gardner2000}, \citet{Madau2000}, \citet{Elbaz2002}, \citet{Metcalfe2003}, \citet{Dole2004}, \citet{Papovich2004}, \citet{Fazio2005}, \citet{Xu2005}, \citet{Frayer2006}, \citet{Levenson2007}, \citet{Thompson2007}, \citet{Levenson2008}, \citet{Bethermin2010}, \citet{Berta2010}, \citet{Keenan2010} and \citet{Voyer2011}. The upper limits shown are from \citet{Dube1979}, \citet{Martin1991}, \citet{Mattila1991}, \citet{Kashlinsky1996} and \citet{Edelstein2000}. Also shown are large-scale anisotropy measurements from \citet{Penin2012} and \citet{Zemkov2014}.}
         \label{Contour}
   \end{figure*}

The energy range has to take into account the redshift dependency in Eq. (3) since the interaction of the $\gamma$-ray and the EBL photons can happen in any point between the Earth and the source. The range is between [$(1+z)^{2}E_{min},E_{max}$], corresponding to a wavelength range of the EBL where the interaction with the $\gamma$-ray can take place along the entire path between the source and the Earth.  In Fig. \ref{Contour} we show the contours from the statistical + systematic uncertainty of the EBL flux density, derived scaling up the EBL template model by \citet{Dominguez2011} at redshift $z=0$ . The wavelength coverage is in the so-called cosmic optical background (COB) part of the EBL, where we found the peak flux density $\lambda F_{\lambda}=12.27_{-2.29}^{+2.75}$ nW m$^{-2}$ sr$^{-1}$ at 1.4 $\mu$m, systematics included.

\section{Conclusions}
\label{sec:conclus}

We have reported the observation of the extraordinary outburst from 1ES 1011+496 observed by MAGIC from February 6th to March 7th 2014 where the flux reached a level $\sim$ 14 times the observed flux at the time of the discovery of the source in 2007. 
The spectrum of 1ES 1011+496 during this flare displays little intrinsic curvature over $>1$ order of magnitude in energy, which makes this an ideal observation for constraining the EBL. Although the source showed a high flux variability during the observed period, no significant change of the spectral shape was observed during the flare, which allowed us to use the average observed spectrum in the search for an imprint on it of the EBL-induced absorption of $\gamma$ rays. Such EBL imprint can be seen in the fit residuals of the best-fit achieved under the no-EBL assumption (Fig. \ref{Residuals}, left). In the approach that we followed along this work, for the description of the intrinsic spectrum at VHE we restricted ourselves to smooth concave functions which have shown in the past to provide good fits to BL Lac spectra whenever the expected EBL absorption was negligible. Under this assumption, the best-fit EBL density at $\lambda =1.4 \mu$m is $F_{\lambda}=12.27_{-2.29}^{+2.75}$ nW m$^{-2}$ sr$^{-1}$, which ranks among the strongest EBL density constraints obtained from VHE data of a single source. 

\begin{acknowledgements}
We would like to thank
the Instituto de Astrof\'{\i}sica de Canarias
for the excellent working conditions
at the Observatorio del Roque de los Muchachos in La Palma.
The financial support of the German BMBF and MPG,
the Italian INFN and INAF,
the Swiss National Fund SNF,
the ERDF under the Spanish MINECO, and
the Japanese JSPS and MEXT
is gratefully acknowledged.
This work was also supported
by the Centro de Excelencia Severo Ochoa SEV-2012-0234, CPAN CSD2007-00042, and MultiDark CSD2009-00064 projects of the Spanish Consolider-Ingenio 2010 programme,
by grant 268740 of the Academy of Finland,
by the Croatian Science Foundation (HrZZ) Project 09/176 and the University of Rijeka Project 13.12.1.3.02,
by the DFG Collaborative Research Centers SFB823/C4 and SFB876/C3,
and by the Polish MNiSzW grant 745/N-HESS-MAGIC/2010/0.
We thank the anonymous referee for a thorough review and a very constructive list of remarks that helped improving the quality and clarity of this manuscript.
\end{acknowledgements}

\bibliographystyle{aa}
\bibliography{biblio1ES1011_v2}

\end{document}